# New Interfaces for Musical Expression

**Ivan Poupyrev**
Interaction Lab
Sony CSL
3-14-13 Higashi-Gotanda
Tokyo 141-0022, Japan
poup@csl.sony.co.jp

**Michael J. Lyons**
ATR MIC Labs
2-2 Hikaridai, Seika
Souraku-gun
Kyoto 619-02, Japan
mlyons@mic.atr.co.jp

**Sidney Fels**
Department of ECE
UBC
Vancouver, BC
Canada, V6T 1Z4
ssfels@ece.ubc.ca

**Tina Blaine (Bean)**
School of Computer Science
Carnegie Mellon University
5000 Forbes Avenue
Pittsburgh, PA 15213, USA
bean@cs.cmu.edu

## ABSTRACT
The rapid evolution of electronics, digital media, advanced materials, and other areas of technology, is opening up unprecedented opportunities for musical interface inventors and designers. The possibilities afforded by these new technologies carry with them the challenges of a complex and often confusing array of choices for musical composers and performers. New musical technologies are at least partly responsible for the current explosion of new musical forms, some of which are controversial and challenge traditional definitions of music. Alternative musical controllers, currently the leading edge of the ongoing dialogue between technology and musical culture, involve many of the issues covered at past CHI meetings. This workshop brings together interface experts interested in musical controllers and musicians and composers involved in the development of new musical interfaces.

## INTRODUCTION
This workshop brings together interface experts interested in musical controllers and musicians and composers involved in the development of new musical interfaces, especially alternative controllers, to stimulate exchange with the following aims:

(1) To survey and discuss the current state of control interfaces for musical performance, identify current and promising directions of research and unsolved problems. To focus on the major practical concerns involved in the design of interfaces for musical expression.

(2) To identify major issues involved the interplay between technological change and changes in musical forms.

(3) To identify the ways in which alternate controllers affect the overall creative process from composition to performance and determine what impact this has on musical expression.

(4) To put together the collective working experience and wisdom of the participants in some tangible form, such as strategies for success and a list of the 10 most difficult problems in musical controls.

## MUSICAL INTERFACES: PAST, PRESENT, FUTURE
Music has historically been a meeting point for technology and artistic expression. The design of musical instruments may well have been the first area of technology where careful and systematic interface design played an essential if not the central role. While music has always been a driving force for technological innovation, it is also true that new technologies have opened the way for new forms of musical expression and experimentation. To give a familiar example, the modern piano, and consequently the classical piano repertoire, such as Beethoven concertos, would not be possible without the great improvements in metallurgy at the turn of the 18th century. This allowed the construction of one-piece cast-iron frames that could support the 18-ton string tensions exerted by performers (Saches, 1940).

In the current era, new technologies that can benefit musical expression are appearing at an accelerating pace. The last century, especially in the 1950's and 60's, saw the rise of electronic musical sound synthesis which gave birth to a plethora of new musical forms in both popular and classical or "serious" arenas of electronic music. We can expect that the continuing progress in information technologies will stimulate composers and musicians to experiment with new means of composition and new instruments for performance.

The development of novel sensor interfaces, vision and pattern recognition, virtual and augmented reality, haptic feedback devices and the like are all opening up avenues for new musical adventurers. The field of alternative musical controllers is at a stage somewhat similar to where electronic synthesis was in the 1950's. The basic paradigms are still being explored and there is an explosion of new interfaces, with, so far, little systematic thought about where the field is headed. Because alternative controllers are essentially means of mapping human behavior into musical expression, issues dealt with by interface designers could be very helpful in understanding and clarifying the state of the field.

## GENERAL ISSUES IN MUSICAL CONTROLLERS
(1) The explosion of methods for generating, sequencing, layering and controlling sounds offers a complex and often confusing range of choices to musical explorers. The limits of music are being pushed, and many in the audience ask: is this music? In the context of new interfaces for human musical expression, fundamental questions are raised: what is a composition; what is a musical instrument? In the case of a machine assisted composition or performance are we still listening to music? How do interface issues bear upon these age-old fundamental questions about art and aesthetics in the context of music?

(2) The rapid pace of change of the new technologies used to build new controllers is double edged: there is a growth of exciting new controllers - however these controllers risk becoming technologically obsolete very quickly. Will establishing standards help or hinder musical interface evolution? The MIDI standard is a case in point - though it has become a wide spread standard in electronic music, there is nonetheless controversy about whether its influence is overall positive or negative.

(3) Alternative controllers bring new freedom to musical expression in that the mapping between action and sound-generation can be arbitrarily changed. However, relatively few mappings are intuitive and natural, many do not make use of our physical intuition and are difficult to learn and use. It is therefore important to discuss what features of mappings constitute appropriate musical interface design.

## MUSICAL CONTROLLERS: DESIGN ISSUES

While part of the workshop will aim to stimulate basic inquiry into the impact of interface technology on musical culture, the main body of the workshop will be devoted to a dialogue amongst the participants on the practical matter of how to design good musical interfaces.

(1) To identify criteria for evaluating musical interfaces.

- usability and comprehensibility
- expressiveness
- sensitivity and sophistication
- aesthetics
- hedonics ("does it feel good?")

and other criteria. The aim is to clarify what guidelines are needed to develop interfaces that are worthy of the dedication and practice needed in acquiring skill with a new instrument.

(2) Identify key interface technology developments that offer the most exciting new opportunities for musical expression. For example

- touch sensors
- position, orientation, and motion sensors
- pressure and strain sensors
- gloves and suits
- computer vision and pattern recognition
- tactile feedback

How is sensor input best mapped onto musical sound? Are certain modes of human behavior and motion more suitable for musical expression than others?

(3) To discuss the role of cognitive science and psychology in the design of musical interfaces. What are the factors determining whether an interface is suitable for the creative expression of complex ideas and emotion patterns?

(4) To share collective experience. Each participant should include in their position their experience on designing musical alternative controllers, their advantages and disadvantages in musical performance. Distilling these reports, the group will try to suggest effective design patterns or guidelines.

## CONCLUSIONS

The rapid evolution of electronics, digital media, advanced materials, and other areas of technology, is opening up unprecedented opportunities for musical interface inventors and designers. The possibilities afforded by these new technologies carry with them the challenges of a complex and often confusing array of choices for musical composers and performers. New musical technologies are at least partly responsible for the current explosion of new musical forms, some of which are controversial and challenge traditional definitions of music. Alternative musical controllers, currently the leading edge of the ongoing dialogue between technology and musical culture, involve many of the issues covered at past CHI meetings. This workshop brings together interface experts interested in musical controllers and musicians and composers involved in the development of new musical interfaces.